\documentclass[12pt,letterpaper]{article}
\usepackage[margin=1in]{geometry}
\usepackage{amsmath,amssymb}
\usepackage{graphicx}
\usepackage{hyperref}
\usepackage{listings}
\usepackage{xcolor}
\usepackage{booktabs}
\usepackage{tabularx}
\newcolumntype{U}{>{\small\ttfamily\raggedright\arraybackslash}X}
\usepackage{url}
\usepackage{natbib}
\bibpunct{(}{)}{;}{a}{,}{,}
\usepackage{tikz}
\usetikzlibrary{shapes.geometric, arrows.meta, positioning, fit, backgrounds}
\usepackage{float}

\hypersetup{
    colorlinks=true,
    linkcolor=blue,
    citecolor=blue,
    urlcolor=blue
}

\title{A Model Context Protocol Server for Astrophysical RAG:\\
Unified Access to HI, Dwarf, Globular Cluster, IntZ, and ALPINE\\
Kinematic Corpora with FAISS Semantic Search}

\author{David C.\ Flynn\\
EPS Research, Laurel, MD, USA\\
ORCID: 0000-0002-2768-6650\\
\texttt{davidflynn@eps-research.com}}

\date{August 2026}

\begin{document}

\maketitle

\begin{abstract}
\sloppy
We present the EPS Research Astro-RAG MCP Server v2.3.0, a cross-platform
Model Context Protocol (MCP) implementation providing unified, machine-readable
access to five astrophysical corpora spanning the local universe ($z = 0$) to
$z = 5.68$ (the high-redshift frontier of the ALPINE survey). The server exposes a
human-readable browser query interface, a REST API, and an LLM-native MCP
endpoint, enabling deterministic retrieval, metadata filtering, and structured
analysis across 2,064 objects: galaxies spanning HI rotation curve surveys, dwarf
and irregular systems, and high-redshift kinematic targets, together with Milky
Way globular clusters. Version 2.3.0 introduces FAISS-accelerated
natural-language similarity search using pre-built 384-dimensional MiniLM-L6-v2
vector indexes, enabling corpus-wide semantic queries without fine-tuning or API
keys. We describe the server architecture, unified schema design, FAISS index
construction pipeline, MCP toolset, and cross-epoch use cases including
rotation-curve retrieval, metadata filtering, and semantic similarity
exploration. The server is publicly deployed on HuggingFace Spaces
(\url{https://dflynn5656-astro-rag-mcp.hf.space}), released under the MIT
License, and fully reproducible from Zenodo
(DOI:~\href{https://doi.org/10.5281/zenodo.21154451}{10.5281/zenodo.21154451}).
The full platform is available at
\url{https://github.com/eps-research/rag-corpus-series}.
\end{abstract}

\section{Introduction}

The proliferation of multi-epoch astrophysical survey data has created a
retrieval problem: individual corpora are well-documented within their own
communities but difficult to query programmatically across epochs, surveys, or
object classes without significant custom tooling. A researcher wishing to
compare rotation-curve kinematics at $z = 0$ with morpho-kinematic
classifications at $z \approx 1$ or $z \approx 5$ must navigate heterogeneous
data formats, inconsistent field naming, and separate access infrastructures for
each survey archive.

Large language models (LLMs) offer a path toward natural-language data access,
but their utility depends on structured retrieval infrastructure that can serve
corpus data deterministically and at low latency. The Model Context Protocol \citep{Anthropic2024}, an open specification for LLM-native tool APIs, provides
a standardized interface for this purpose: an MCP-compliant server exposes a
typed toolset that any MCP-compatible AI assistant (Claude, Copilot, and others)
can invoke directly, without bespoke integration code.

We present the EPS Research Astro-RAG MCP Server v2.3.0, built on this
foundation. The server provides unified access to five machine-readable
kinematic corpora assembled under the EPS Research RAG Corpus Series
\citep{Flynn2026platform}: the Unified HI Rotation Curve Corpus v7.0 (438
galaxies), the Dwarf/Irregular HI Corpus v1.0 (129 galaxies), the Milky Way
Globular Cluster Corpus v1.3.2 (174 clusters), the IntZ Kinematic Corpus v1b
(1,292 galaxies, $z \approx 0.4$--2.7), and the High-z ALPINE Kinematic Corpus Z1 (31 galaxies, $z = 4.26$--5.68; \citealt{LeFevre2020,Bethermin2020}). Together these constitute 2,064 objects
with a unified schema, full survey provenance, and citable Zenodo DOIs for each
corpus.

Version 2.3.0 adds FAISS-accelerated semantic search \citep{Johnson2021} over
all five corpora simultaneously, enabling queries such as ``dwarf irregular low
surface brightness'' or ``metal-poor outer halo cluster'' to return semantically
ranked object lists without knowledge of field names or query syntax. This
capability distinguishes the server from programmatic API clients such as
\texttt{astroquery} \citep{Ginsburg2019}, which require database-specific module
selection and formal query syntax.

\section{The EPS Research RAG Corpus Series}

The EPS Research RAG Corpus Series is a unified open-science platform comprising
five machine-readable kinematic corpora, 149 executable Jupyter notebooks, and
supporting tools. All corpora share a common JSON schema with native-typed fields
(float, int, bool, null), embedded survey provenance, and consistent field naming
across epochs. The platform is publicly available at
\url{https://github.com/eps-research/rag-corpus-series} and described in detail
in Flynn (2026) \citep{Flynn2026platform}.

Table~\ref{tab:corpora} summarizes the five corpora served by the MCP server.

\begin{table}[ht]
\centering
\caption{EPS Research RAG corpora served by the MCP server v2.3.0.}
\label{tab:corpora}
\begin{tabularx}{\textwidth}{l c c U l}
\toprule
Corpus & Key & $N$ & Redshift range & Zenodo DOI \\
\midrule
Unified HI Rotation Curve Corpus v7.0 & \texttt{v7}    & 438   & $z \approx 0$      & 10.5281/zenodo.19563417 \\
Dwarf/Irregular HI Corpus v1.0        & \texttt{dwarf} & 129   & $z \approx 0$      & 10.5281/zenodo.20320362 \\
Milky Way GC Corpus v1.3.2            & \texttt{gc}    & 174   & MW                 & 10.5281/zenodo.19907766 \\
IntZ Kinematic Corpus v1b$^{a}$       & \texttt{intz}  & 1,292 & $z \approx 0.4$--2.7 & 10.5281/zenodo.21841382 \\
High-z ALPINE Kinematic Corpus Z1     & \texttt{z1}    & 31    & $z = 4.26$--5.68  & 10.5281/zenodo.21834678 \\
\midrule
\textbf{Total}                        &                & \textbf{2,064} & $z = 0$--5.68 & \\
\bottomrule
\end{tabularx}
\par\vspace{2pt}{\footnotesize $^a$ IntZ $\omega$ values are unavailable (\texttt{omega\_available = false}); KROSS boundary points were template-derived rather than independently measured rotation-curve rings (August 2026; see CORRECTIONS.md at \url{https://github.com/eps-research/rag-corpus-series}).}
\end{table}

\section{Server Architecture}

\subsection{Technology Stack}

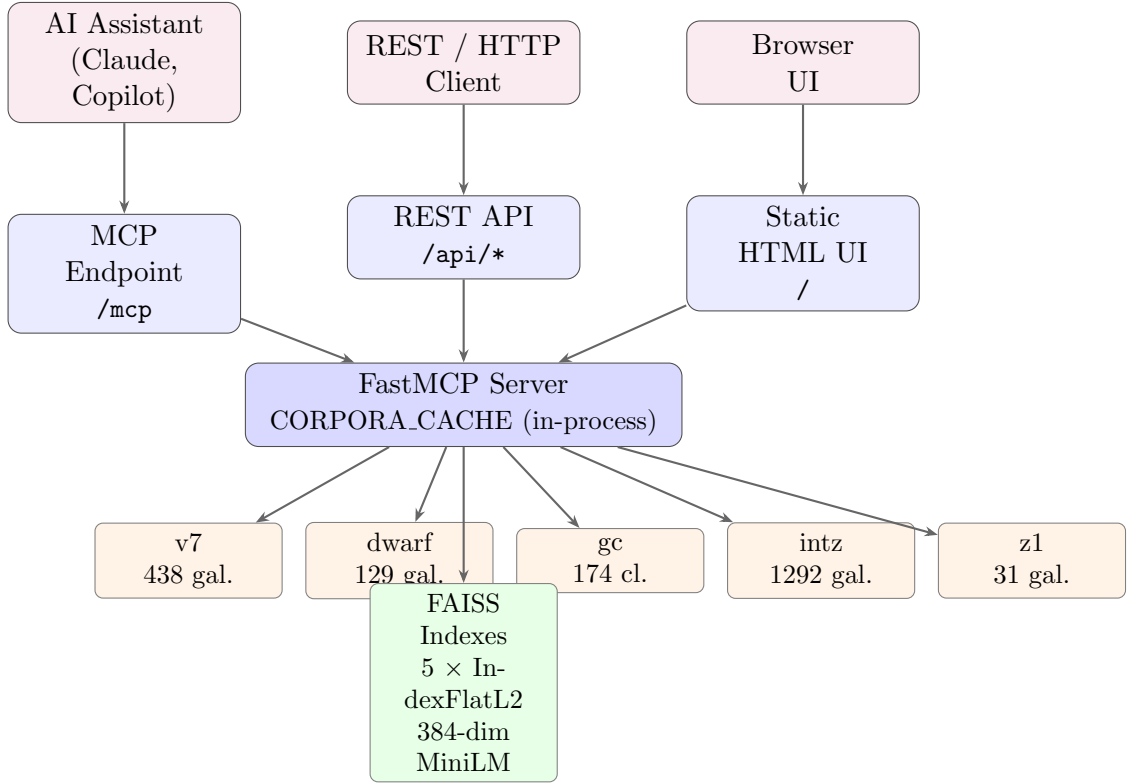
\begin{figure}[ht!]
\centering
\tikzset{
  block/.style={rectangle, rounded corners=5pt, draw=black!70, fill=blue!8,
                text width=2.8cm, minimum height=1.1cm, align=center, font=\small},
  corpus/.style={rectangle, rounded corners=3pt, draw=black!50, fill=orange!10,
                text width=2.2cm, minimum height=0.8cm, align=center, font=\footnotesize},
  faiss/.style={rectangle, rounded corners=3pt, draw=black!50, fill=green!10,
                text width=2.2cm, minimum height=0.8cm, align=center, font=\footnotesize},
  arr/.style={-{Stealth[length=5pt]}, thick, draw=black!60},
  lbl/.style={font=\footnotesize\itshape, text=black!55}
}
\begin{tikzpicture}[node distance=0.7cm and 1.4cm]

\node[block, fill=purple!8] (mcp_client) {AI Assistant\\(Claude, Copilot)};
\node[block, fill=purple!8, right=of mcp_client] (rest_client) {REST / HTTP\\Client};
\node[block, fill=purple!8, right=of rest_client] (browser) {Browser\\UI};

\node[block, below=1.2cm of mcp_client] (mcp_ep) {MCP\\Endpoint\\{\footnotesize\texttt{/mcp}}};
\node[block, below=1.2cm of rest_client] (rest_ep) {REST API\\{\footnotesize\texttt{/api/*}}};
\node[block, below=1.2cm of browser] (ui_ep) {Static\\HTML UI\\{\footnotesize\texttt{/}}};

\node[block, fill=blue!15, below=1.1cm of rest_ep, text width=5.5cm] (core) 
  {FastMCP Server\\{\footnotesize CORPORA\_CACHE (in-process)}};

\node[corpus, below left=1.0cm and -0.5cm of core] (v7) {v7\\438 gal.};
\node[corpus, right=0.3cm of v7] (dwarf) {dwarf\\129 gal.};
\node[corpus, right=0.3cm of dwarf] (gc) {gc\\174 cl.};
\node[corpus, right=0.3cm of gc] (intz) {intz\\1292 gal.};
\node[corpus, right=0.3cm of intz] (z1) {z1\\31 gal.};

\node[faiss, below=1.8cm of core] (faiss) {FAISS Indexes\\{\footnotesize 5 × IndexFlatL2}\\{\footnotesize 384-dim MiniLM}};

\draw[arr] (mcp_client) -- (mcp_ep);
\draw[arr] (rest_client) -- (rest_ep);
\draw[arr] (browser) -- (ui_ep);

\draw[arr] (mcp_ep) -- (core);
\draw[arr] (rest_ep) -- (core);
\draw[arr] (ui_ep) -- (core);

\draw[arr] (core) -- (v7);
\draw[arr] (core) -- (dwarf);
\draw[arr] (core) -- (gc);
\draw[arr] (core) -- (intz);
\draw[arr] (core) -- (z1);

\draw[arr] (core) -- (faiss);

\end{tikzpicture}
\caption{EPS Research Astro-RAG MCP Server v2.3.0 architecture. Three concurrent
interfaces (MCP endpoint, REST API, browser UI) are served by a single FastMCP
process. Corpus data is held in an in-process cache (\texttt{CORPORA\_CACHE});
FAISS indexes are loaded at startup and queried by the \texttt{semantic\_search}
tool. All five corpora share a unified JSON schema.}
\label{fig:arch}
\end{figure}

The MCP server is implemented in Python using FastMCP \citep{FastMCP2024}, a
high-level framework for building MCP-compliant servers. Figure~\ref{fig:arch}
shows the server architecture. The server exposes
three concurrent interfaces over a single process:

\begin{enumerate}
    \item \textbf{MCP endpoint} (\texttt{/mcp}): LLM-native tool API implementing
    the Anthropic MCP specification with Server-Sent Events (SSE) and Streamable
    HTTP transports.
    \item \textbf{REST API} (\texttt{/api/*}): FastAPI-backed HTTP endpoints
    with full OpenAPI/Swagger documentation at \texttt{/docs}, providing
    programmatic access without an MCP client.
    \item \textbf{Browser UI} (\texttt{/}): A single-page HTML interface
    providing point-and-click corpus queries, semantic search, and JSON/text
    result export.
\end{enumerate}

Corpus data is loaded from JSONL files at startup and held in an in-process
cache (\texttt{CORPORA\_CACHE}), with asynchronous HTTP fetching via
\texttt{httpx}. Cold-start latency is approximately 30--90 seconds on the
HuggingFace Spaces free tier, depending on whether FAISS indexes require
downloading from GitHub; subsequent requests are served from the in-process
cache at sub-millisecond latency.

\subsection{Deployment}

The server is deployed on HuggingFace Spaces
(\url{https://dflynn5656-astro-rag-mcp.hf.space}) using a Docker container.
A UptimeRobot keepalive ping every 5 minutes prevents Space hibernation,
maintaining effectively zero cold-start latency for regular users. A
cross-platform launcher (\texttt{launch.py}) with GUI progress popups and
automatic dependency management supports local deployment on Windows 11 and
Ubuntu.

\section{MCP Toolset}

The server exposes seven MCP tools, each returning a structured JSON object
with a \texttt{human\_text} field for direct LLM consumption and a \texttt{data}
field for programmatic use. Table~\ref{tab:tools} summarizes the toolset.

\begin{table}[ht]
\centering
\caption{MCP tools exposed by the server. All tools accept a \texttt{corpus}
parameter with values \texttt{v7}, \texttt{dwarf}, \texttt{gc}, \texttt{intz},
\texttt{z1}. Return types are JSON objects with \texttt{success}, \texttt{tool},
\texttt{data}, and \texttt{human\_text} fields.}
\label{tab:tools}
\begin{tabularx}{\textwidth}{l U}
\toprule
Tool & Description \\
\midrule
\texttt{list\_corpora}      & Returns metadata, record counts, redshift ranges, and Zenodo DOIs for all five corpora. \\
\texttt{list\_objects}      & Returns paginated object ID lists with optional survey and quality-tier filters. \\
\texttt{get\_object}        & Retrieves the full JSON record for any object by ID, including all survey-merged fields and provenance. \\
\texttt{search\_metadata}   & Case-insensitive substring search over any metadata field across all records in a corpus. \\
\texttt{filter\_objects}    & Numeric range filter on any field (e.g., \texttt{distance\_mpc}, \texttt{z\_spec}); supports \texttt{omega\_ready\_only} flag. \\
\texttt{get\_corpus\_schema} & Returns the field schema and column definitions for any corpus. \\
\texttt{semantic\_search}   & Natural-language similarity search using FAISS vector indexes; returns top-$k$ ranked objects. \\
\bottomrule
\end{tabularx}
\end{table}

Tool signatures are typed and documented in the MCP manifest, allowing
AI assistants to select the appropriate tool automatically from a natural-language
request. For example, the query ``find galaxies with rotation velocity above
150~km~s$^{-1}$'' routes to \texttt{filter\_objects} with \texttt{field =
vrot\_max\_kms}, \texttt{min = 150}, while ``find galaxies similar to DDO~154''
routes to \texttt{semantic\_search} with \texttt{corpus = v7}, \texttt{query =
DDO 154 dwarf irregular}.

\section{FAISS Semantic Search}

\subsection{Index Construction}

FAISS \citep{Johnson2021} indexes are pre-built offline for each corpus using
the \texttt{all-MiniLM-L6-v2} model from the \texttt{sentence-transformers} library \citep{Reimers2019},
which produces 384-dimensional dense embeddings optimized for semantic
similarity. For each object record, a curated summary string is constructed
from key metadata fields rather than the full JSON record, to focus the
embedding on scientifically meaningful content. For example, a v7 galaxy
record produces a string of the form:

\begin{lstlisting}
Galaxy NGC3198. survey THINGS. morphology SBc (Sc). distance 13.8 Mpc.
inclination 72.0 deg. max rotation velocity 150.0 km/s. quality tier 1.
omega ready True. telescope VLA.
\end{lstlisting}

Field selection is corpus-specific: GC records embed metallicity, galactic
coordinates, dynamical mass, and core-collapse flag; intz records embed
redshift, stellar mass, star-formation rate, and kinematic class; Z1 records
embed tracer line, redshift, rotation classification, and ALPINE survey
identifiers.

Indexes use \texttt{faiss.IndexFlatL2} (exact nearest-neighbor search), which
is appropriate for corpus sizes of order $10^3$ objects and guarantees
reproducible, deterministic results. The 15 index files (five \texttt{.faiss}
binaries, five \texttt{\_ids.json} ID mappings, five \texttt{\_texts.json}
embedding text files) are deposited at Zenodo
(DOI:~\href{https://doi.org/10.5281/zenodo.21147895}{10.5281/zenodo.21147895})
and hosted on GitHub for automatic download at server startup.

\subsection{Query Flow}

At runtime, a \texttt{semantic\_search} call (1) encodes the query string with
the cached MiniLM model, (2) performs an \texttt{index.search} against the
pre-loaded FAISS index for the requested corpus, and (3) returns the top-$k$
object IDs with L2 similarity scores. Query latency is dominated by the
embedding step ($\sim$10--50~ms on CPU), with FAISS search adding negligible
overhead for $N \leq 2{,}064$.

\subsection{Validation}

Table~\ref{tab:semantic} shows representative queries and top-3 returned
objects for each corpus, demonstrating semantic coherence across object
classes and redshift regimes.

\begin{table}[ht]
\centering
\caption{Representative semantic search queries and top-3 results across all
five corpora. Queries are natural-language strings with no field-name knowledge
required.}
\label{tab:semantic}
\begin{tabularx}{\textwidth}{l l U}
\toprule
Corpus & Query & Top-3 returned objects \\
\midrule
\texttt{v7}    & dwarf irregular low mass          & CVnIdwA, DDO~210, IC~1613 \\
\texttt{gc}    & metal poor outer halo cluster     & ESO-SC06, 2MS-GC02$^a$, ESO~93-8 \\
\texttt{dwarf} & Sculptor group low rotation       & UGCA442, ESO349-G031, ESO245-G005 \\
\texttt{intz}  & high redshift rotating disk KROSS & S-CFHT-NBJ-1478, C-HiZ\_z1\_111, C-HiZ\_z1\_112 \\
\texttt{z1}    & confirmed rotator ALPINE CII      & VC5110377875, CG32, J0817 \\
\bottomrule
\end{tabularx}
\end{table}
{\footnotesize $^a$ The corpus uses the abbreviated designation 2MS-GC02; the standard literature form is 2MASS-GC02 \citep{Ivanov2005}.}

\section{Cross-Epoch Use Cases}

\subsection{Rotation Curve Retrieval}

An LLM assistant connected via MCP can retrieve a full rotation curve record
with a single \texttt{get\_object} call. The returned JSON includes all
survey-merged fields: inclination, position angle, distance, kinematic model,
$V_\mathrm{rot}$ profile, and provenance. Cross-validation between the v7 and
dwarf corpora is enabled by consistent field naming: \texttt{vrot\_max\_kms},
\texttt{inc\_deg}, \texttt{distance\_mpc}, and \texttt{kinematic\_model} are
present in both. The following Python example demonstrates direct REST access:

\begin{lstlisting}[language=Python]
import requests

r = requests.get(
    "https://dflynn5656-astro-rag-mcp.hf.space/api/get_object",
    params={"corpus": "v7", "object_id": "NGC3198"}
)
record = r.json()["data"]["record"]
print(record["vrot_max_kms"], record["distance_mpc"])
# 150.0  13.8
\end{lstlisting}

\subsection{Metadata Filtering}

The \texttt{filter\_objects} tool enables range queries on any numeric field.
For example, filtering \texttt{v7} on \texttt{distance\_mpc} between 1 and
10~Mpc returns 119 of 438 galaxies; filtering \texttt{intz} on \texttt{z\_spec}
between 0.5 and 1.0 returns 822 of 1,292 objects. These queries complete in
under 100~ms and require no knowledge of database schema beyond the field name.

\begin{lstlisting}[language=Python]
# Filter v7 galaxies within 10 Mpc
r = requests.get(
    "https://dflynn5656-astro-rag-mcp.hf.space/api/filter_objects",
    params={"corpus": "v7", "field": "distance_mpc", "min": 1, "max": 10}
)
data = r.json()["data"]
print(data["n_matched"], "galaxies within 10 Mpc")
# 119 galaxies between 1 and 10 Mpc

# Filter intz galaxies at z=0.5-1.0
r = requests.get(
    "https://dflynn5656-astro-rag-mcp.hf.space/api/filter_objects",
    params={"corpus": "intz", "field": "z_spec", "min": 0.5, "max": 1.0}
)
print(r.json()["data"]["n_matched"], "intz galaxies at z=0.5-1.0")
# 822 intz galaxies at z=0.5-1.0
\end{lstlisting}

\subsection{Semantic Similarity Exploration}

The \texttt{semantic\_search} tool enables discovery queries that would be
impossible with exact-match retrieval. The query ``galaxies the same size as
DDO~161'' returns DDO161, DDO168, DDO170, DDO\_133, and DDO\_52 — a
physically coherent set of small late-type dwarfs. The query ``core-collapsed
high concentration globular cluster'' returns clusters with confirmed
post-core-collapse morphology in the GC corpus.

\begin{lstlisting}[language=Python]
# Semantic search across v7 corpus
r = requests.get(
    "https://dflynn5656-astro-rag-mcp.hf.space/api/semantic_search",
    params={"corpus": "v7", "query": "dwarf irregular low mass", "top_k": 5}
)
results = r.json()["data"]["results"]
for obj in results:
    print(obj["id"], f"score={obj['score']:.3f}")
# CVnIdwA   score=1.216
# DDO_210   score=1.218
# IC_1613   score=1.219
# DDO_154   score=1.221
# DDO_168   score=1.223

# Cross-corpus: find similar objects in dwarf corpus
r = requests.get(
    "https://dflynn5656-astro-rag-mcp.hf.space/api/semantic_search",
    params={"corpus": "dwarf", "query": "Sculptor group low rotation", "top_k": 3}
)
print([x["id"] for x in r.json()["data"]["results"]])
# ['UGCA442', 'ESO349-G031', 'ESO245-G005']
\end{lstlisting}

\section{Relation to Existing Infrastructure}

\texttt{astroquery} \citep{Ginsburg2019} provides programmatic access to
specific databases via formal APIs and is the standard tool for targeted
catalog queries. The Astro-RAG MCP server is complementary: it targets
pre-assembled, schema-unified corpora rather than live database queries,
and is designed for LLM-native consumption rather than script-based workflows.
The two tools are best used together: \texttt{astroquery} for fresh
database queries, the MCP server for structured retrieval and semantic
exploration of the assembled corpora.

The Virtual Observatory (VO) stack (TAP, ADQL, VOTable) provides powerful
structured query capabilities over VO-compliant archives. The MCP server
does not replicate VO infrastructure but instead provides a complementary
LLM-native interface optimized for natural-language access and cross-epoch
corpus exploration.

No existing public MCP server provides unified kinematic access across the
full $z = 0$--5.68 redshift range with semantic search capability. To our
knowledge, this is the first MCP-compliant astrophysical data server.

\section{Discussion}

\subsection{LLM-Native Data Access}

The MCP specification enables a qualitatively new mode of astrophysical data
access: an AI assistant with the server connected can answer questions such as
``list all KROSS galaxies with stellar mass above $10^{10}$~M$_\odot$ and
velocity dispersion above 80~km~s$^{-1}$'' by composing \texttt{filter\_objects}
and \texttt{search\_metadata} calls, without the user needing to know the field
names \texttt{log\_mass\_msun} or \texttt{sigma0\_kms}. This lowers the barrier
to corpus exploration for researchers unfamiliar with the schema.

\subsection{Reproducibility}

All components are versioned and deposited at Zenodo with citable DOIs: the
server software (this work), the five corpora (separate deposits), and the FAISS
indexes. The server source is available at GitHub. Any result obtained via the
MCP server can be reproduced from the Zenodo deposits without network access to
HuggingFace.

\subsection{Limitations}

The FAISS indexes use L2 distance over MiniLM embeddings, which captures
semantic similarity in natural-language descriptions but does not encode
physical distance in parameter space. Two galaxies at the same distance in
embedding space may differ substantially in, e.g., inclination or kinematic
model quality. Numeric similarity queries are better served by
\texttt{filter\_objects}; semantic search is most useful for discovery and
reconnaissance. The intz corpus contains records with missing
\texttt{identifiers.name} fields that return numeric index IDs rather than
object names in semantic search results; this will be corrected in a future
corpus version.

\section{Conclusions}

We have presented the EPS Research Astro-RAG MCP Server v2.3.0, a
cross-platform MCP implementation providing unified LLM-native access to five
astrophysical kinematic corpora spanning $z = 0$ to $z = 5.68$. The server
exposes deterministic retrieval, metadata filtering, and FAISS-accelerated
semantic search over 2,064 objects via a standardized MCP toolset, a REST API,
and a browser query interface. It is publicly deployed, free to use, and fully
reproducible from Zenodo deposits.

The platform represents, to our knowledge, the first MCP-compliant
astrophysical data server with cross-epoch kinematic coverage and semantic
search capability. It is available at
\url{https://github.com/eps-research/rag-corpus-series} and
\url{https://dflynn5656-astro-rag-mcp.hf.space}.

\section*{Acknowledgments}

The author thanks the developers of FastMCP, FAISS, sentence-transformers,
FastAPI, and HuggingFace Spaces for the open-source infrastructure on which
this server is built. The author acknowledges the survey teams behind SPARC \citep{Lelli2016}, THINGS \citep{Walter2008}, LITTLE THINGS \citep{Hunter2012}, WHISP, LVHIS \citep{Koribalski2018}, KROSS, KMOS3D, and ALPINE \citep{LeFevre2020,Bethermin2020} for making their kinematic data publicly accessible. This work was carried out at EPS
Research, an independent research organization, without institutional
affiliation or grant funding.

\section*{Data Availability}

The MCP server software is available at
\href{https://doi.org/10.5281/zenodo.21154451}{Zenodo DOI:~10.5281/zenodo.21154451}
and \url{https://github.com/eps-research/rag-corpus-series}.
The FAISS semantic search indexes are available at
\href{https://doi.org/10.5281/zenodo.21147895}{Zenodo DOI:~10.5281/zenodo.21147895}.
The five corpora are available at their respective Zenodo DOIs listed in
Table~\ref{tab:corpora}. The live server endpoint is
\url{https://dflynn5656-astro-rag-mcp.hf.space}.


\begin{thebibliography}{99}

\bibitem[Anthropic(2024)]{Anthropic2024}
Anthropic 2024, \textit{Model Context Protocol Specification},
\url{https://modelcontextprotocol.io}

\bibitem[Flynn(2026a)]{Flynn2026platform}
Flynn, D.~C.\ 2026a, \textit{The EPS Research Astro-RAG Platform},
arXiv:2605.30384

\bibitem[Flynn(2026b)]{Flynn2026hi}
Flynn, D.~C., \& Cannaliato, J.\ 2026b, \textit{Unified HI Rotation Curve
Corpus v7.0}, Zenodo, DOI:~10.5281/zenodo.19563417

\bibitem[Flynn(2026c)]{Flynn2026dwarf}
Flynn, D.~C.\ 2026c, \textit{Dwarf/Irregular HI Corpus v1.0}, Zenodo,
DOI:~10.5281/zenodo.20320362

\bibitem[Flynn(2026d)]{Flynn2026gc}
Flynn, D.~C.\ 2026d, \textit{Milky Way Globular Cluster Corpus v1.3.2}, Zenodo,
DOI:~10.5281/zenodo.19907766

\bibitem[Flynn(2026e)]{Flynn2026intz}
Flynn, D.~C.\ 2026e, \textit{IntZ Kinematic Corpus v1b (corrected; $\omega$ withdrawn)}, Zenodo,
DOI:~10.5281/zenodo.21841382

\bibitem[Flynn(2026f)]{Flynn2026z1}
Flynn, D.~C.\ 2026f, \textit{High-z ALPINE Kinematic Corpus Z1 (v3; $\omega$ corrected August 2026)}, Zenodo,
DOI:~10.5281/zenodo.21834678

\bibitem[Flynn(2026g)]{Flynn2026faiss}
Flynn, D.~C.\ 2026g, \textit{EPS Research RAG Corpus Series --- FAISS Semantic
Search Indexes v1.0}, Zenodo, DOI:~10.5281/zenodo.21147895

\bibitem[Le F{\`e}vre et al.(2020)]{LeFevre2020}
Le F{\`e}vre, O., Béthermin, M., Faisst, A., et al.\ 2020,
\textit{Astronomy \& Astrophysics}, 643, A1

\bibitem[Béthermin et al.(2020)]{Bethermin2020}
B{\'e}thermin, M., Fudamoto, Y., Ginolfi, M., et al.\ 2020,
\textit{Astronomy \& Astrophysics}, 643, A2

\bibitem[FastMCP(2024)]{FastMCP2024}
FastMCP Development Team 2024, \textit{FastMCP: The Fast, Pythonic Way to
Build MCP Servers}, \url{https://github.com/jlowin/fastmcp}

\bibitem[Ginsburg et al.(2019)]{Ginsburg2019}
Ginsburg, A., Sip{\H o}cz, B.~M., Brasseur, C.~E., et al.\ 2019,
\textit{The Astronomical Journal}, 157, 98

\bibitem[Hunter et al.(2012)]{Hunter2012}
Hunter, D.~A., Ficut-Vicas, D., Ashley, T., et al.\ 2012,
\textit{The Astronomical Journal}, 144, 134

\bibitem[Ivanov et al.(2005)]{Ivanov2005}
Ivanov, V.~D., Kurtev, R., \& Borissova, J.\ 2005,
\textit{Astronomy \& Astrophysics}, 435, 949

\bibitem[Johnson et al.(2021)]{Johnson2021}
Johnson, J., Douze, M., \& J{\'e}gou, H.\ 2021,
\textit{IEEE Transactions on Big Data}, 7, 535

\bibitem[Koribalski et al.(2018)]{Koribalski2018}
Koribalski, B.~S., Wang, J., Kamphuis, P., et al.\ 2018,
\textit{Monthly Notices of the Royal Astronomical Society}, 478, 1611

\bibitem[Lelli et al.(2016)]{Lelli2016}
Lelli, F., McGaugh, S.~S., \& Schombert, J.~M.\ 2016,
\textit{The Astronomical Journal}, 152, 157

\bibitem[Reimers \& Gurevych(2019)]{Reimers2019}
Reimers, N., \& Gurevych, I.\ 2019, \textit{Proceedings of the 2019
Conference on Empirical Methods in Natural Language Processing},
\url{https://arxiv.org/abs/1908.10084}

\bibitem[Walter et al.(2008)]{Walter2008}
Walter, F., Brinks, E., de Blok, W.~J.~G., et al.\ 2008,
\textit{The Astronomical Journal}, 136, 2563

\end{thebibliography}
\end{document}